\documentclass[6pt,twocolumn,showpacs,preprintnumbers,amsmath,amssymb,pra]{revtex4}

\usepackage{xcolor}
\usepackage{times}
\usepackage[T1]{fontenc}
\usepackage{microtype}
\usepackage{graphicx}
\usepackage{dcolumn}
\usepackage{bm}
\usepackage{cancel}
\usepackage{soul}

\newcommand{\dint}[2]{\mathrm{d}^{#1}{#2}}

\newcommand{\ii}{\mathrm{i}}
\makeatletter
\def\imod#1{\allowbreak\mkern10mu {\operator@font mod}\,\,#1 }
\makeatother

\begin{document}


\title{Time-dependent second Born calculations\\for model atoms and molecules in strong laser fields}
\author{K. Balzer}
\email{balzer@theo-physik.uni-kiel.de}
\author{S. Bauch}
\author{M. Bonitz}
\affiliation{Institut f\"ur Theoretische Physik und Astrophysik, Christian-Albrechts-Universit\"at Kiel, Leibnizstrasse 15, 24098 Kiel, Germany}

\date{\today}

\begin{abstract}
Using the finite-element discrete variable representation of the nonequilibrium Green's function (NEGF) we extend previous work [K.~Balzer et al., Phys. Rev. A \textbf{81}, 022510 (2010)] to nonequilibrium situations and compute---from the two-time Schwinger-Keldysh-Kadanoff-Baym equations---the response of the helium atom and the heteronuclear molecule lithium hydride to laser fields in the uv and xuv regime. In particular, by comparing the one-electron density and the dipole moment to time-dependent Hartree-Fock results on one hand and the full solution of the time-dependent Schr\"odinger equation on the other hand, we demonstrate that the time-dependent second Born approximation carries valuable information about electron-electron correlation effects. Also, we outline an efficient distributed memory concept which enables a parallel and well scalable algorithm for computing the NEGF in the two-time domain.
\end{abstract}



\maketitle

\section{\label{sec:intro}Introduction}
Owing to binary interactions, electrons inside and between atoms and molecules form a highly sophisticated quantum many-particle system. Already two-electron atoms show rich structure~\cite{tanner00}, e.g.~revealed by the presence of autoionization resonances above the lowest ionization threshold. Therefore, when out of equilibrium, and, especially, when exposed to intense laser fields, atoms and molecules comprise very complex electron dynamics which occur at sub-femtosecond time scales and include nonlinear excitations, electron scattering and ionization~\cite{krausz09}. Novel techniques, such as attosecond streaking (using coherent, ultrashort laser pulses)~\cite{itatani02}, allow one to capture and trace this dynamics experimentally and give, e.g., time- and angle-resolved indication of single, double or multi ionization processes. One of the most prominent examples, in this context, are attosecond (two-color) pump-probe scenarios, e.g.~\cite{uiberacker07}, including ionization that follows an electron shake up~\cite{bauch10}. Thereby, laser sources allow for the preparation of specially selected atomic or molecular initial states.

In all aforementioned processes, electron-electron (e-e) correlations play a substantial role. Thus, consistent methods beyond the single-active electron picture are required to describe electronic structure and dynamics. However, theory largely depends on good approximate solutions to the many-electron problem and, particularly, time-resolved investigations including correlations
in sub-femtosecond regimes are still in their infancy.

In this paper, we, in an ab-initio fashion, apply generalized kinetic equations, the Schwinger-Keldysh-Kadanoff-Baym equations (SKKBE)~\cite{martin59,keldysh64,kadanoff62}, to compute the electron motion in model atoms and molecules exposed to laser fields. In contrast to time-dependent density functional theory, NEGFs, thereby, offer a systematic approach to e-e correlations, where the residual interactions beyond the time-dependent Hartree-Fock (TDHF) frame can be incorporated in the SKKBE by self-consistent (i.e.~$\phi$-derivable) and conserving approximations of the interaction kernel~\cite{kadanoff62}. In comparison to (approximate) wave function based methods, such as the time-dependent Kohn-Sham orbital approach~\cite{bauer97,ruggenthaler09} and its multiconfigurational variants~\cite{zanghellini04,hochstuhl10_jpcs}, NEGF theory does not provide immediate access to all observables as it is premised on a two-time generalization of the reduced (one-particle) density matrix. However, all one- and---due to the two-time dependencies---also the two-particle quantities can, in principle, be obtained at the same level of self-consistency~\cite{kadanoff62,vanleeuwen06}. A key for applying NEGFs to nonequilibrium laser-atom interaction is an efficient, well adapted representation of the Green's function. In Ref.~\cite{balzer10_pra}, we have shown that a suitable concept is given by the finite-element discrete variable representation (FE-DVR). Here, we extend this work to nonequilibrium.

It is the goal of this paper to prove feasibility and adequacy of two-time NEGF calculations along with spatially extended atomic hamiltonians. To this end, we consider $N$ nonrelativistic electrons which move in a time-dependent potential $V(x,t)$ and interact via a pair potential $U(x-x')$. Sec.~\ref{sec:theory} reviews the equations of motion for the associated one-particle NEGF in FE-DVR representation, where e-e correlations are treated in time-dependent second Born approximation (TD2ndB). In Sec.~\ref{sec:results}, we present results for ultraviolet-field (uv) induced electron dynamics in case of the helium~\cite{pindzola91,grobe93,liu99,dahlen01,lein00} and lithium hydride model~\cite{tempel09} including, respectively, two and four electrons. Thereby, we compare TDHF and TD2ndB to full solutions of the time-dependent Schr\"odinger equation (TDSE).

\section{\label{sec:theory}Theory}
We begin the theory section by briefly describing the grid-based representation of the one-particle NEGF $G(1,1')=-\ii\langle\hat{T}_{\cal C}{\psi}(1){\psi}^\dagger(1')\rangle$, $1=(x,t,\sigma)$, in coordinate space by means of a FE-DVR basis---for a more detailed description the reader is referred to Refs.~\cite{balzer10_pra} and \cite{balzer10_jpcs}. Thereafter, Sec.~\ref{subsec:eom} recapitulates the respective equations of motion, and Sec.~\ref{subsec:mpi} discusses our concept for parallel distributed-memory computation of $G(1,1')$ using the message passing interface (MPI).
Atomic units (a.u.) are used throughout, $m=|e|=\hbar=4\pi\varepsilon_0=1$.

\subsection{\label{subsec:representation}Representation of $G(1,1')$}
We represent the NEGF in one space dimension on an interval $[0,x_0]$ which is partitioned into $n_e$ finite elements $[x^i,x^{i+1}]$, $i\in\{0,1,\ldots,n_e-1\}$. The FE boundaries are at the same time generalized Gauss-Lobatto (GGL) points $x^i=x^i_0$ and $x^{i+1}=x^i_{n_g}$ of a GGL quadrature of order $n_g$ with abscissa points $x^i_m$ and weights $w_m^i$, $m\in\{0,1,\ldots,n_g-1\}$, for details see e.g.~\cite{balzer10_pra} and references therein. These points and weights define local DVR basis sets~\cite{rescigno00}, which can be superposed such that the total FE-DVR space is spanned by
\begin{align}
\label{chi}
 \chi^i_m(x)=
\left\{
\begin{array}{cc}
\displaystyle\frac{f^i_{n_g-1}(x)+f^{i+1}_0(x)}{[w^i_{n_g-1}+w^{i+1}_0]^{1/2}}\;,  &m=0\\
&\\
f^i_m(x) [w^i_m]^{-1/2}\;, &m\neq0\;,\\
\end{array}
\right.
\end{align}
where $f^i_m(x)$ are Lobatto shape functions~\cite{rescigno00}. All $\chi_0^i(x)$---denoted bridge functions---ensure the NEGF being continuous in coordinate space, cf.~Ref.~\cite{balzer10_pra}. Moreover, the basis has dimension $n_b=n_e n_g-1$ and is \textit{GGL orthonormal}~\cite{expl_gglorthonormal}. Using Eq.~(\ref{chi}), the one-particle NEGF $G(1,1')=\theta(t-t')\,G^>(1,1')+\theta(t'-t)\,G^<(1,1')$ in FE-DVR representation is expanded as
\begin{align}
\label{gdef}
 G(xt,x't')&=\sum_{ii'}\sum_{mm'}\chi^i_m(x)\,\chi^{i'}_{m'}(x')\,g_{mm'}^{ii'}(t,t')\;,&\\
&\hspace{7.75pc}g_{mm'}^{ii'}(t,t')\in\mathbb{C}\;,\nonumber
\end{align}
where we imply a spin-restricted ansatz, i.e.~$G$ does not explicitly depend on the spin-projection $\sigma$. However, the spin-degeneracy $\xi\in\{1,2\}$ is incorporated in the self-energies, i.e. $\Sigma(1,1')=\Sigma_\xi(xt,x't')$, the representation of which is as $G$ in Eq.~(\ref{gdef}) with time-dependent coefficients $ \Sigma_{\xi,mm'}^{ii'}(t,t')\in\mathbb{C}$.

\subsection{\label{subsec:eom}Equations of motion}
In FE-DVR representation, the second-quantized form of the considered $N$-electron Hamiltonian reads as
\begin{align}
\label{ham2}
 \hat{H}&=\sum_{ii'}\sum_{mm'}h_{mm'}^{ii'}(t)[c^{i}_{m}]^\dagger c_{m'}^{i'}&\\
&+\sum_{i_1\ldots i_4}\sum_{m_1\ldots m_4}u_{m_1m_2m_3m_4}^{i_1i_2i_3i_4}[c^{i_1}_{m_1}]^\dagger [c^{i_3}_{m_3}]^\dagger c_{m_2}^{i_2} c_{m_4}^{i_4}\;,\nonumber
\end{align}
where the operator $c_m^i$ ($[c_m^i]^\dagger$) creates (annihilates) an electron in the local state $\chi_m^i(x)$. In particular, it holds
\begin{align}
\label{1pi}
 h_{mm'}^{ii'}(t)&=\int_0^{x_0}\!\!\!\dint{}{x}\,\chi_m^i(x)\left\{-\frac{1}{2}\nabla^2+V(x,t)\right\}\chi_{m'}^{i'}(x)\nonumber\\
&=t_{mm'}^{ii'}+\delta_{mm'}^{ii'}\tilde{v}_{m}^{i}(t)\;,
\end{align}
for the one-electron integrals and
\begin{align}
\label{2pi}
 u_{m_1m_2m_3m_4}^{i_1i_2i_3i_4}&=\int_0^{x_0}\!\!\!\dint{}{x}\!\int_0^{x_0}\!\!\!\dint{}{x'}\,\chi_{m_1}^{i_1}(x)\chi_{m_3}^{i_3}(x')\nonumber\\
&\hspace{5.75pc}\times U(x-x')\chi_{m_2}^{i_2}(x)\chi_{m_4}^{i_4}(x')&\nonumber\\
&=\delta^{i_1i_2}_{m_1m_2}\delta^{i_3i_4}_{m_3m_4}\tilde{u}_{m_1m_3}^{i_1i_3}\;,
\end{align}
for the two-electron integrals, where $\delta_{ii'}^{mm'}=\delta_{ii'}\delta_{mm'}$. Explicit expressions for the kinetic energy $t_{mm'}^{ii'}$, the time-dependent potential energy $\tilde{v}_m^i(t)$ and the interaction matrix $\tilde{u}_{mm'}^{ii'}$ are given in Ref.~\cite{balzer10_pra}. We note, that the equality in the last lines of Eqs.~(\ref{1pi}) and (\ref{2pi}), respectively, is only true when the GGL quadrature rule is being applied.

Concerning Eq.~(\ref{ham2}), in FE-DVR picture, the equations of motion for
$g_{mm'}^{ii'}(t,t')=-\ii\langle\hat{T}_{\cal C} c_m^i(t) [c_{m'}^{i'}(t')]^\dagger\rangle$
are the SKKBE transformed into the matrix equation
\begin{align}
\label{skkbe}
\{\ii&\partial_t\delta_{i\bar{i}}^{m\bar{m}}-h_{m\bar{m}}^{i\bar{i}}(t)\}g_{\bar{m}m'}^{\bar{i}i'}(t,t')&\nonumber\\
&=\delta_{\cal C}(t-t')\delta_{ii'}^{mm'}+\int_{\cal C}\!\dint{}{\bar{t}}\,\Sigma_{\xi,m\bar{m}}^{i\bar{i}}[g,u](t,\bar{t})g_{\bar{m}m'}^{\bar{i}i'}(\bar{t},t')\;,
\end{align}
where $\cal C$ denotes the complex Schwinger-Keldysh contour and additional summation is implied over $\bar{i}$ and $\bar{m}$. The self-energy $\Sigma_{\xi, mm'}^{ii'}[g,u](t,t')$ depends functionally upon the matrices $g$ and $u$ and, in Hartree-Fock and second Born approximation, is defined in the Appendix. Due to the simple structure of Eq.~(\ref{2pi}) they are easily evaluated in FE-DVR representation.

The SKKBE~(\ref{skkbe}) represents the basic equation to be solved, in Sec.~\ref{sec:results}, for atomic and molecular model systems. Thereby, to resolve the correlated dynamics (TD2ndB), it has to be propagated into the two-time plane ${\cal P}=[0,t_f]\times[0,t_f]$ which scales quadratically with final propagation time $t_f$. To do so, we follow the scheme described in Refs.~\cite{dahlen07_prl,stan09}.

\begin{figure}
\includegraphics[width=0.375\textwidth]{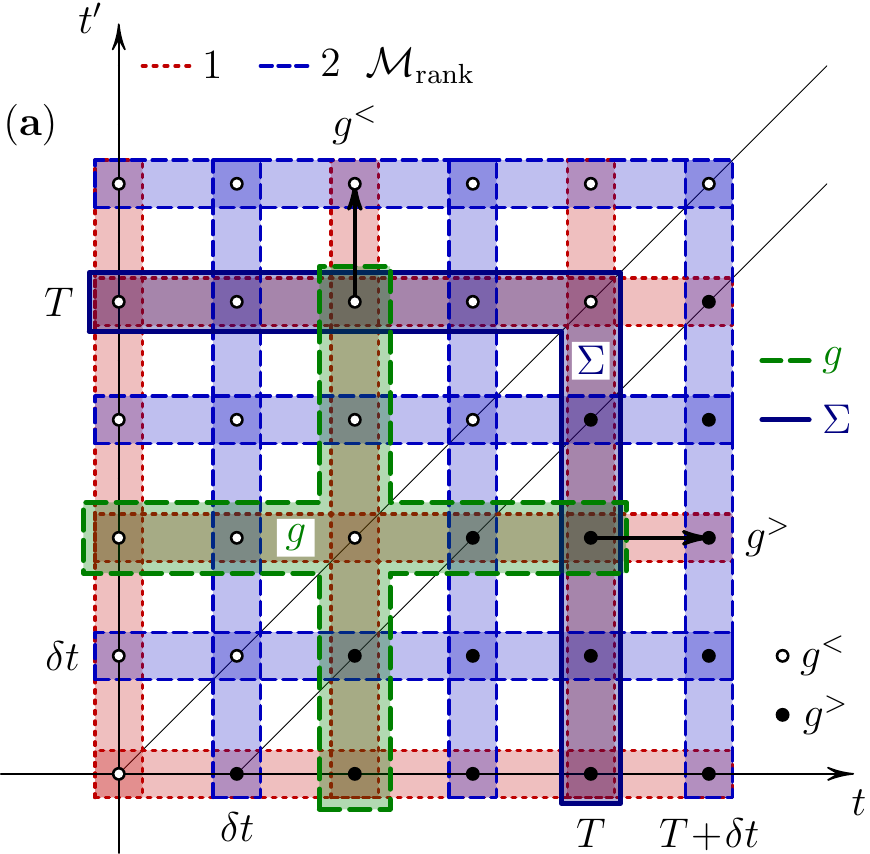}
\includegraphics[width=0.375\textwidth]{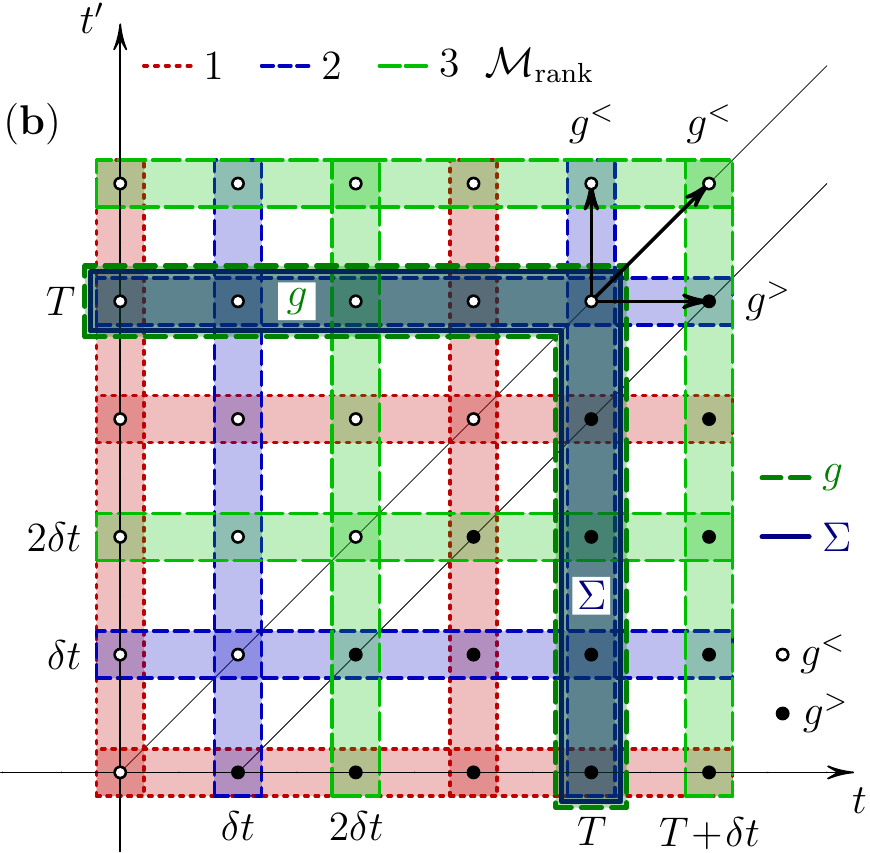}
 \caption{(Color online) Distributed-memory concept in the case of (a) two MPI processes, (b) three MPI processes. The different domains with dotted, dashed and long-dashed border (thin lines) mark main memory for $g^\gtrless(t,t')$ allocated by distinct processes (ranks). The extent of the memory kernel [the r.h.s.~of the SKKBE] for the indicated time steps $T\rightarrow T+\delta T$ (cf.~the arrows) is given by the thick dashed line, while the required self-energies $\Sigma^\gtrless(t,t')$ are bordered with a thick solid line. Note, that the relevant history is completely known to the process that performs the respective time step. In addition, the propagation of the greater and lesser Green's function (extending the upper and lower triangle) can be performed by a single process.
}\label{fig.dm}
\end{figure}

\subsection{\label{subsec:mpi}Code parallelization and performance}
While the TDHF part of the self-energy, being of first order in $u$, can be absorbed into $h_{m\bar{m}}^{i\bar{i}}(t)$ of Eq.~(\ref{skkbe}), higher-order contributions are connected to e-e correlations and lead to a memory kernel [r.h.s.~of the SKKBE] which, generally, is not smooth. This means, that, during each single time step $T\rightarrow T+\delta t$ within ${\cal P}$, integrals over the complete dynamics history have to be performed~\cite{stan09}. Consequently, for each point $(t,t')\in{\cal P}$ an $n_b^2$-dimensional array has to be allocated containing the spatial part of the NEGF. For symmetry reasons, the storage of the lesser (greater) correlation function $g^<$ ($g^>$) can be constrained to an upper (lower) triangle, cf.~the open (closed) circles in Fig.~\ref{fig.dm}~(a) or (b). Hence, the total amount of data evaluates to ${\cal D}=16 n_t^2 n_b^2$ bytes, where $n_t=t_f/\delta t$ and double precision is presumed. In typical one-dimensional [atomic or molecular] problems this can easily involve few terabytes of data~\cite{expl_data}.

Such memory (RAM) requirements are conceptually beyond standard shared memory setups. Therefore, we have developed a distributed memory concept~\cite{garny10} based on MPI~\cite{gropp99}. The main memory allocation for $g^\gtrless(t,t')$ is embedded as follows: Depending on the number of processes available (${\cal M}_\mathrm{size}$), the two-time plane ${\cal P}$ is partitioned into a series of different vertical \textit{and} horizontal blocks (arrays) attributed to distinct MPI processes labeled as ${\cal M}_\mathrm{rank}$ in Fig.~\ref{fig.dm}. More precisely, any $g_{mm'}^{\gtrless,ii'}(t,t')$ is simultaneously kept in the memory of process $(t \delta t^{-1})\imod{ {\cal M}_\mathrm{size}}+1$ and in the memory of process $(t'\delta t^{-1})\imod{ {\cal M}_\mathrm{size}}+1$. The situation is illustrated in Fig.~\ref{fig.dm}~(a) for the case of two processes, Fig.~\ref{fig.dm}~(b) refers to ${\cal M}_\mathrm{size}=3$. In general, we arrive at a (multi-colored) chessboard-like pattern which has a ''primitive cell'' of side length $({\cal M}_\mathrm{size}-1)\delta t$. As a consequence, the total NEGF is effectively stored twice, i.e.~${\cal D}_\mathrm{eff}=2 {\cal D}$. However, doing so ascertains that the amount of MPI communication is minimized---this is another critical issue concerning efficient parallel two-time NEGF calculations. In particular, with such memory distribution, no Green's function at all needs to be exchanged with other processes! Exemplarily, the area bordered by the thick dashed line in Fig.~\ref{fig.dm}~(a) indicates that, for a selected time step to be performed by process ${\cal M}_\mathrm{rank}=1$ (cf.~the arrows), the memory [interaction] kernel extends exclusively on local storage domains. The same holds for diagonal time steps, cf.~Fig.~\ref{fig.dm}~(b). Further, note that the propagation of $g^<$ and $g^>$ can be performed in sequence but bunched in a single process. As a result, only the actual self-energy matrices $\Sigma_{mm'}^{<,ii'}(t,T)$ and $\Sigma_{mm'}^{>,ii'}(T,t')$ with $t,t'\leq T$ (see the areas edged by thick solid lines in Fig.~\ref{fig.dm}) need to be broadcasted and made known to all processes, though they can be computed strictly locally.

\begin{figure}[b]
\includegraphics[width=0.48\textwidth]{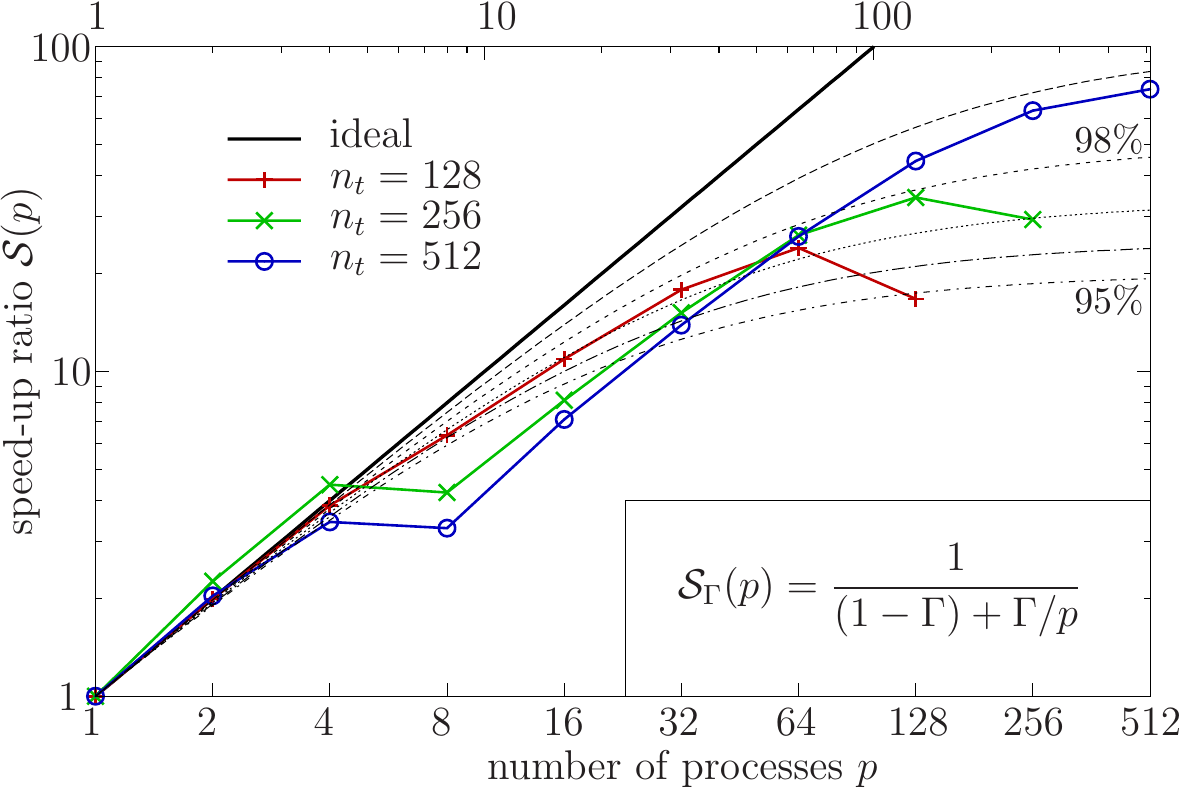}
 \caption{(Color online) Speed-up ratio ${\cal S}(p)=T_{1}/T_{p}$~\cite{expl_speedup} for parallel propagation of $G(1,1')$ with up to $512$ MPI processes for three different propagation times $t_f=n_t \delta t$ where $\delta t=0.025$. As physical system, we have considered the one-dimensional He model (see Sec.~\ref{sec:results}) with $n_b=23$ FE-DVR basis functions. While the shortest calculations (plus symbols, red) have been performed on the ''xe'' nodes of the HLRN batch system~\cite{expl_hlrn}, all other calculations have been carried out by the ''ice1'' nodes. We note, that the performance drop between $4$ and $8$ processes is caused by internal architecture differences. The thin dashed, dotted and dashed-dotted lines indicate a degree of parallelization of $\Gamma=95$\% to $99$\% according to Amdahl's law ${\cal S}_\Gamma(p)$.
}\label{fig.performance}
\end{figure}

In conclusion, such a distributed memory concept allows for an efficient time-stepping algorithm and, at the same time, enables large cluster computations. Regarding the scalability, NEGF calculations concerning the one-dimensional helium atom (cf.~Sec.~\ref{sec:results}) with up to $512$ MPI processes have achieved a typical degree of parallelization of more than $95$\%. Fig.~\ref{fig.performance} shows the corresponding scaling behavior for test calculations of different numbers of processes $p$ and length $n_t$. Finally, we mention that the memory distribution scheme is easily extended to cover also the mixed Green's functions~\cite{stan09}, which account for the self-consistent dynamics of initial correlations.

\begin{figure*}
\includegraphics[width=0.3295\textwidth]{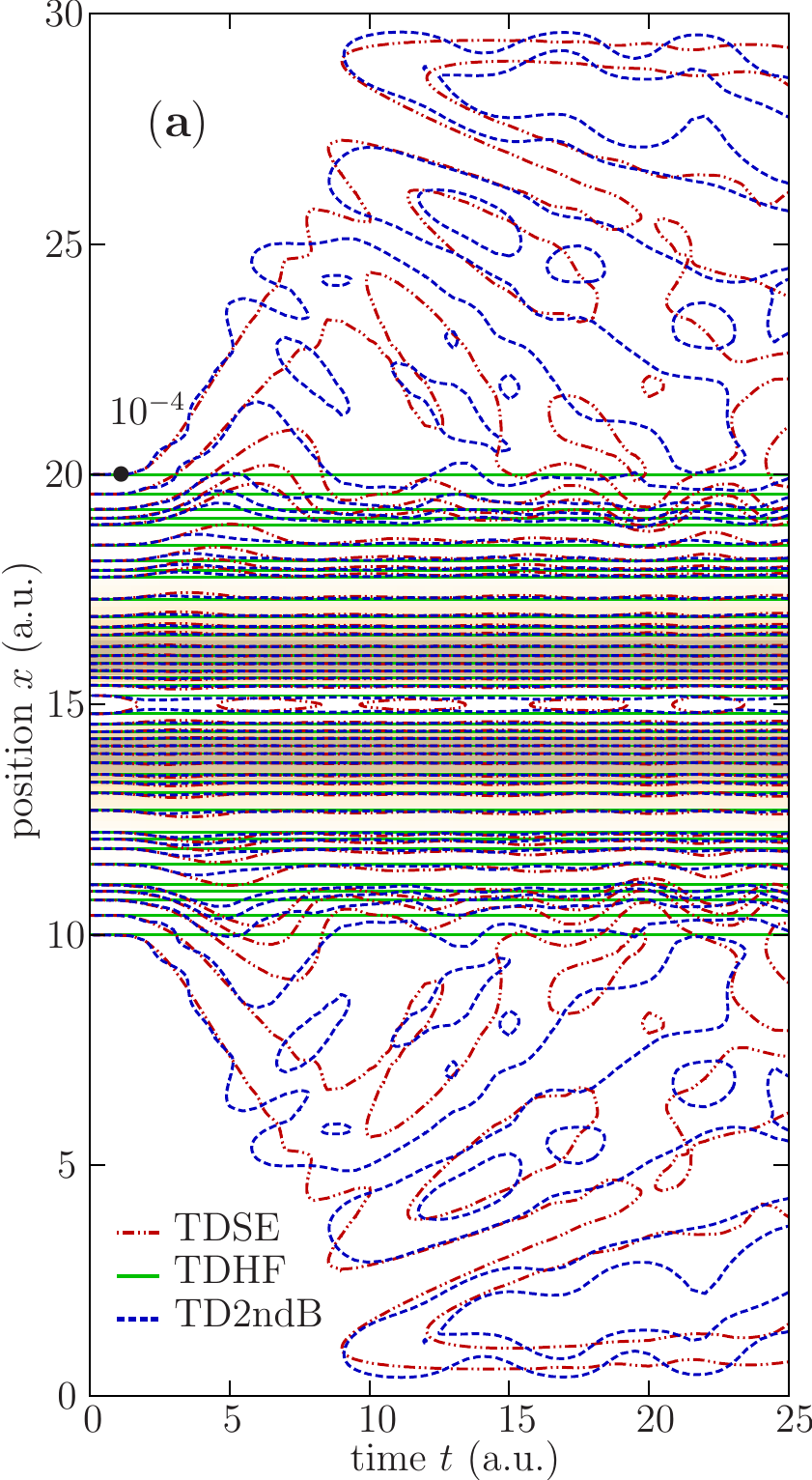}
\includegraphics[width=0.3295\textwidth]{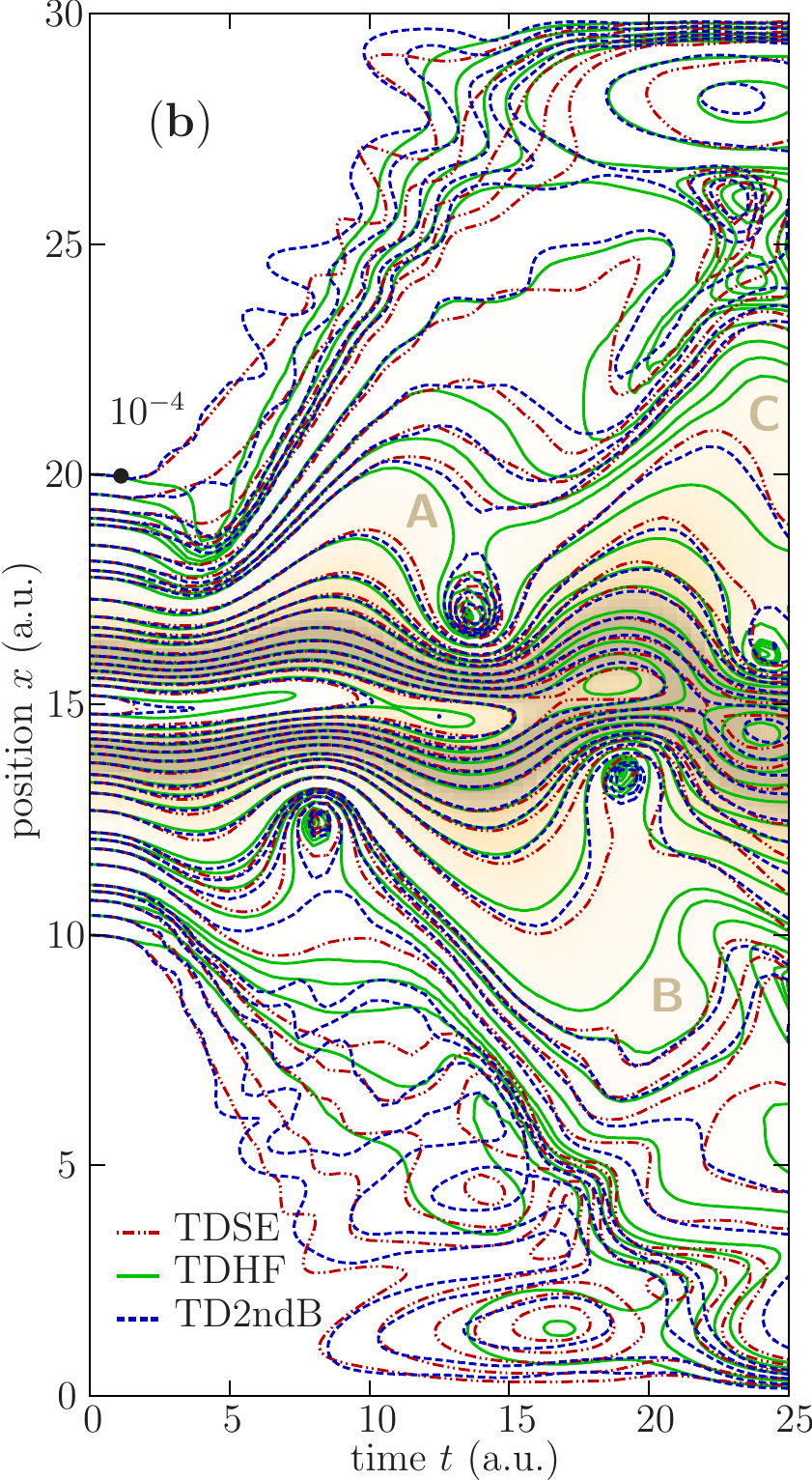}
\includegraphics[width=0.3295\textwidth]{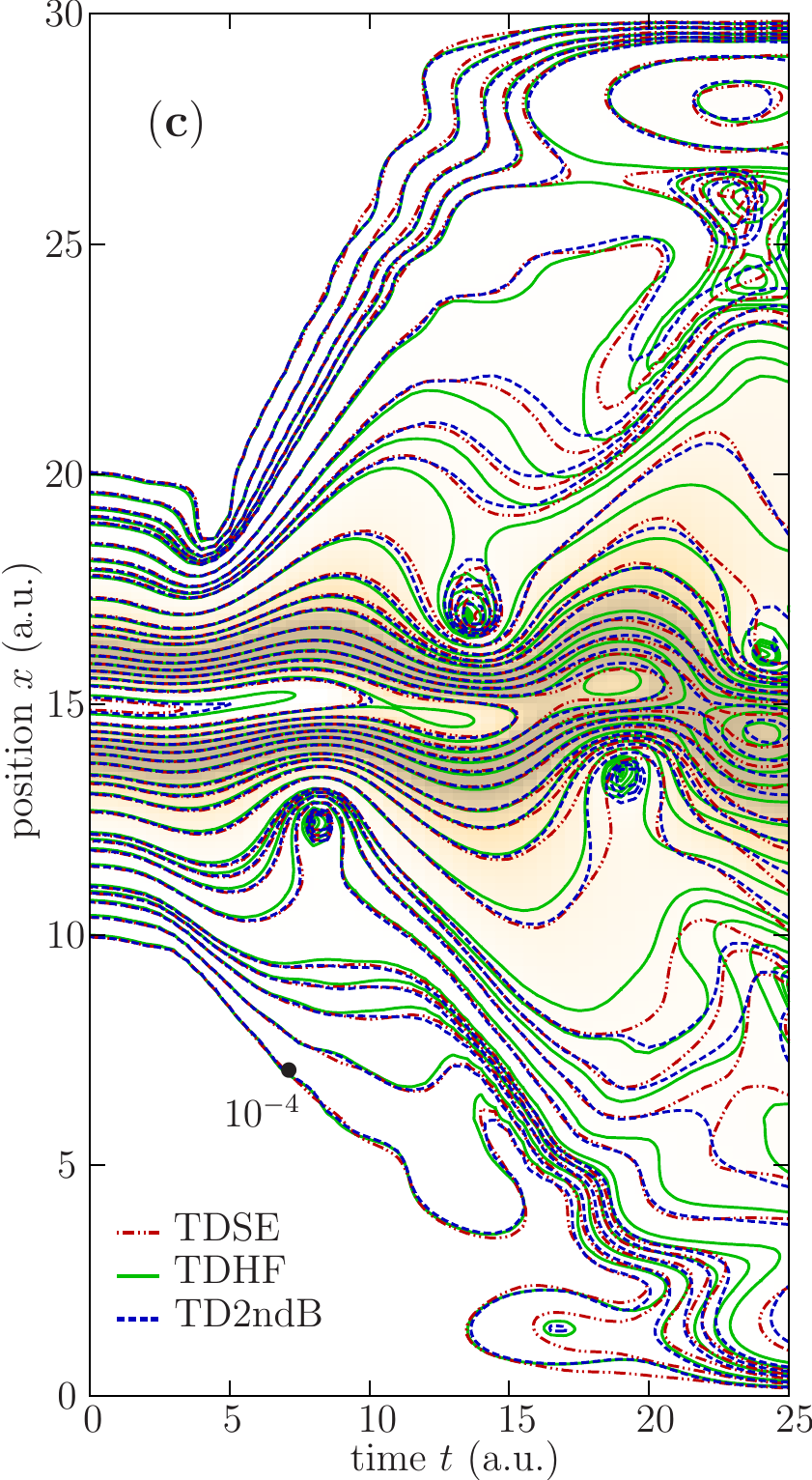}
 \caption{(Color online) Contour plot (logarithmic) showing the time-dependent one-electron density $\langle\hat{n}\rangle(x,t)$ for 1D He in Hartree-Fock (solid, green) and second Born approximation (dashed, blue). The dashed-dot-dotted (red) lines reveal TDSE solutions. (a) Time evolution of the He atom initially prepared in the Hartree-Fock ground state and no uv field being switched on, (b) short-time response of the Hartree-Fock initial state to a permanent uv field ($E_0=0.1$, $\omega_0=0.54$), (c) permanent uv field as in (b) but with the He atom initially prepared in the self-consistent ground state. The contour lines cover a density range from $10^{-4}$ to $0.5$~a.u.~with four contours in each decimal power between $10^{-4}$ and $0.1$, and contours between $0.1$ and $0.5$ each $0.05$~a.u.}\label{fig.1dhedensity}
\end{figure*}

\begin{figure}[t]
\includegraphics[width=0.48\textwidth]{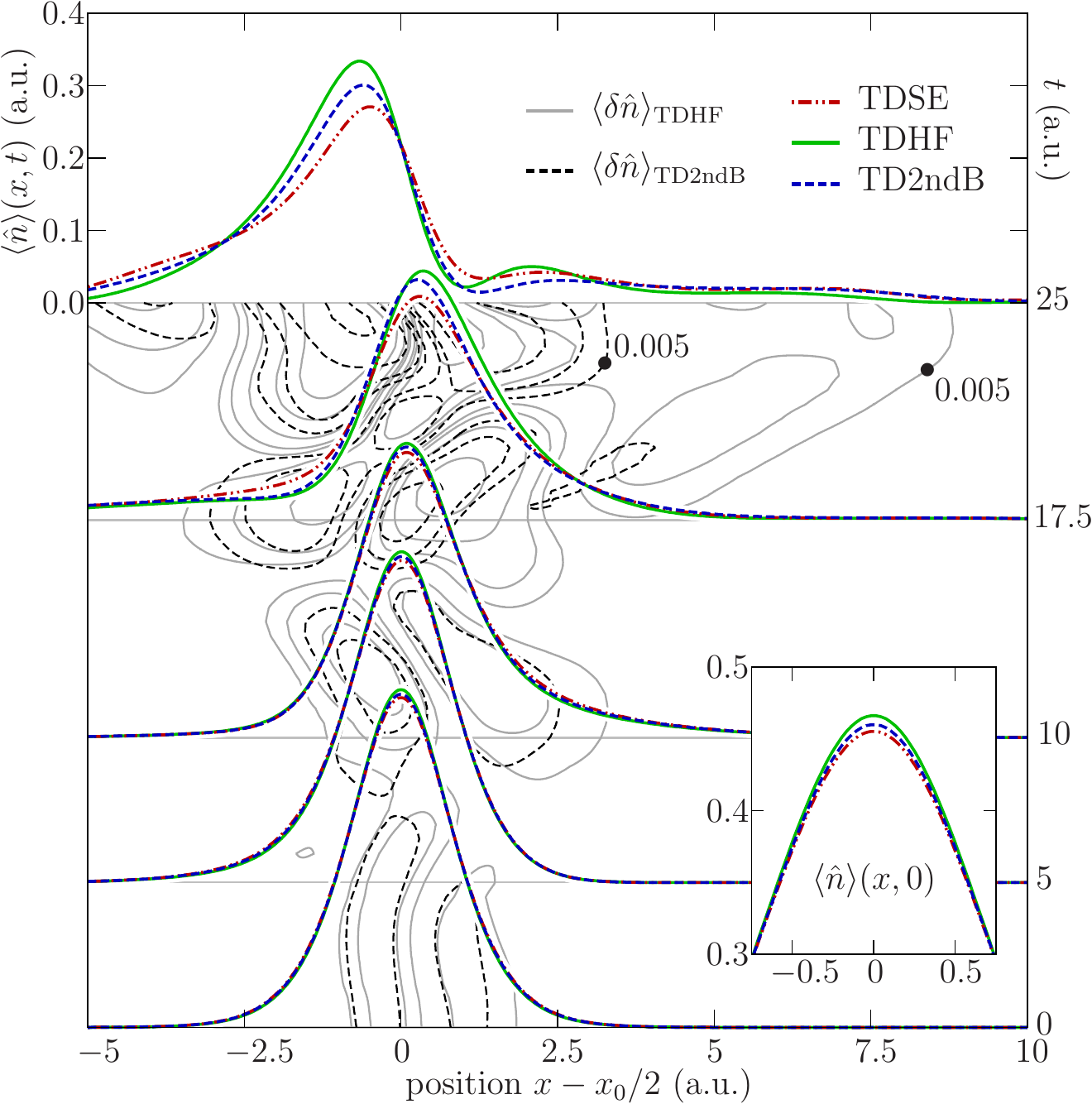}
 \caption{(Color online) Nonlogarithmic plot of the 1D He density $\langle\hat{n}\rangle(x,t)$ for different times $t$ corresponding to Fig.~\ref{fig.1dhedensity}~(c). The inset shows the self-consistent (initial) ground state density at $t=0$. Further, the contour lines indicate the absolute of the density difference $\langle\delta\hat{n}\rangle_\gamma$ of TDHF (solid) and TD2ndB (dashed), respectively, to TDSE at values of $0.005$, $0.01$, $0.02$, $0.04$ and $0.06$.
}\label{fig.1dhenonlogdensity}
\end{figure}

\section{\label{sec:results}Results and discussion}
To illustrate the method, we consider $N$ electrons moving in a one-dimensional model atom or molecule, where the potential in Eq.~(\ref{1pi}) plus a permanent laser field in dipole approximation is given by 
\begin{align}
\label{tdp}
V(x,t)=&-\sum_{m=1}^{M}Z_m U(x-x_{0,m})&\\
       &+ E_0 \cos(\omega_0 t) \theta(t) x\;,\nonumber
\end{align}
with atomic numbers $Z_m$, $M$ nuclei positions $x_{0,m}$, a soft-Coulomb electron-nucleus or, respectively, electron-electron interaction $U(x)=[x^2+1]^{-1/2}$, and an electric field strength (frequency) $E_0$ ($\omega_0$). For $M=1$, the potential $V(x,t)$ describes the core potential of a single atom superposed by the laser field, in particular for $Z_1=2$ we model the helium atom in one spatial dimension (1D He).

\subsection{\label{subsec:he}Helium}
\subsubsection{\label{subsubsec:he1}Short-time density response}
The neutral 1D He model containing two electrons prepared in the singlet ground state has total energy $-2.2383$ (exact). In second Born approximation, the ground state energy is $-2.2334$ accounting for $65$\% of e-e correlations (Hartree-Fock: $-2.2242$), cf.~Ref.~\cite{balzer10_pra}. We now excite this system by uv laser light of intensity $E_0=0.1$ ($3.5\cdot 10^{14}$ W/cm$^2$) and, from the SKKBE (\ref{skkbe}), resolve the time-dependent one-electron density $\langle \hat{n}\rangle(x,t)=-\ii G^<(xt,x't)$ within a simulation box of width $x_0=30$~a.u.~and up to $49$ FE-DVR basis functions. Thereby, the frequency $\omega_0=0.54$ ($84$ nm) is chosen such that the perturbation is [in all cases] near-resonant to the first excited singlet state, cf.~Fig.~\ref{fig.domega}. The short-time density response of the 1D helium atom is shown in Fig.~\ref{fig.1dhedensity} in form of contour plots. Fig.~\ref{fig.1dhedensity}~(a) shows the dynamics in absence of the laser field, $E_0=0$. Here, in TDHF approximation (solid line), the initial state remains an eigenstate of $\hat{H}$ for all times $t$. However, if the Hartree-Fock ground state evolves under influence of e-e correlations, see the dashed (dashed-dot-dotted) TD2ndB (TDSE) curves, collective oscillations in $\langle \hat{n}\rangle(x,t)$ are initiated. These small time-dependent features are, in particular, well characterized in second Born approximation.

In Fig.~\ref{fig.1dhedensity}~(b), laser light is switched on instantaneously at $t=0$, and, as in Fig.~\ref{fig.1dhedensity}~(a), the He atom is prepared in the Hartree-Fock ground state. The external field drives the system out of equilibrium leading to strong density deformations which reveal that both electrons are oscillating in the uv field. Thereby, many time-dependent details in $\langle \hat{n}\rangle(x,t)$ [at high density around $x=15$ and low density] which are not captured in TDHF are well resolved in the TD2ndB calculation, see e.g.~the domains labeled as A, B and C in Fig.~\ref{fig.1dhedensity}~(b). Also, in the TD2ndB and TDSE densities, the oscillatory behavior of Fig.~\ref{fig.1dhedensity}~(a) is superimposed. Moreover, due to the finite simulation box, reflections occur at the interval boundaries.

Fig.~\ref{fig.1dhedensity}~(c) shows the density response after preparation of (correlated) self-consistent initial states, i.e.~propagation of the NEGF in second Born approximation in particular involves mixed Green's functions $G^{\lceil}(xt,x't')$ ($G^{\rceil}(xt,x't')$) with $t$ ($t'$) on the imaginary branch of the contour ${\cal C}$, e.g.~\cite{stan09}. These can be omitted in case of an uncorrelated TDHF calculation. Comparing approximate to exact calculations, we see that TD2ndB performs fairly well and complies essentially better with TDSE than with TDHF. This is observed at high density around $x=15$ as well as for moderate- and low-density domains, where $\langle \hat{n}\rangle(x,t)$ is spatially more extended in TD2ndB approximation. In addition, the oscillatory behavior of (a) and (b) vanishes due to the self-consistency. To support the quality of TD2ndB, Fig.~\ref{fig.1dhenonlogdensity} shows nonlogarithmic snapshots of the electronic density close to the nucleus at different times $t$ and also quantifies the difference $\langle\delta\hat{n}\rangle_{\gamma}=|\langle\hat{n}\rangle_{\gamma} - \langle\hat{n}\rangle_\mathrm{TDSE}|$ of the approximate to the exact density (solid and dashed contour lines). In total, the time-dependent density profile is more spread out due to e-e correlations, and, in contrast to TDHF, the error $\langle\delta\hat{n}\rangle_\mathrm{TD2ndB}$ does not exceed values of $0.05$. Moreover, at specific times during the initial stage of uv excitation, at~$t\approx8$ and $16$~a.u., the deviation to the exact result becomes very small.

\begin{figure}[b]
\includegraphics[width=0.48\textwidth]{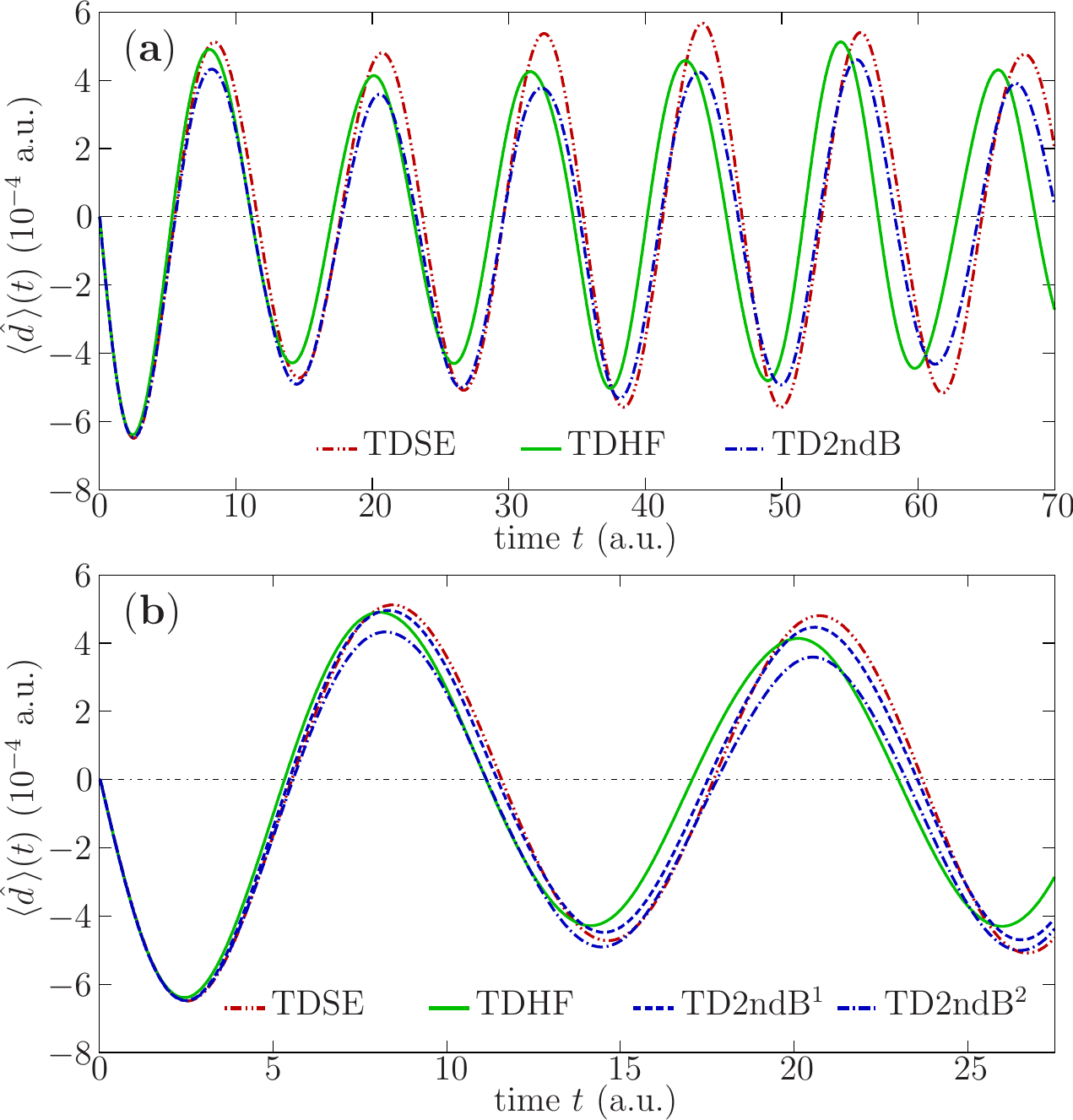}
 \caption{(Color online) Time-dependent dipole moment $\langle\hat{d}\,\rangle(t)$ for the 1D helium atom. (a) TDHF (solid) and TD2ndB approximation (dashed-dotted) versus the exact TDSE result (dashed-dot-dotted). Note that the time-evolution in case of TD2ndB starts from the Hartree-Fock ground state. (b) Comparison of the short-time behavior of $\langle\hat{d}\,\rangle(t)$ for different initial states: TD2ndB$^2$ (dashed-dotted) refers to the second Born approximation with Hartree-Fock initial state, whereas the initial state for TD2ndB$^1$ (dashed) has been computed self-consistently.
}\label{fig.dt}
\end{figure}

\subsubsection{\label{subsubsec:he2}Time-dependent dipole moment}
>From the NEGF we also can access the time-dependent dipole moment (DM) $\langle\hat{d}\,\rangle(t)=\ii\int_0^{x_0}\!\dint{}{x}G^<(xt,xt)$. For the helium atom, the DM is displayed in Fig.~\ref{fig.dt}~(a) implying the following: (i)~To suppress the effect of box boundary reflections we have enlarged the simulation box to $x_0=70$ a.u.~and (ii)~have modified the perturbation to a spectrally broad dipole kick $E_0 \cos(\omega t) \theta(t) x\rightarrow E_0\delta(t)x$ with intensity $E_0=0.01$ ($3.5\cdot 10^{12}$ W/cm$^2$). The latter change allows for a direct computation of dipole spectra via Fourier transform of the DM time-series. Again, in Fig.~\ref{fig.dt}~(a), we compare TDHF and TD2ndB to the exact solution of the TDSE. In second Born approximation, see the dashed-dotted line, the time-dependent DM well approaches the TDSE result with a general shift to a larger [main] oscillation period compared to TDHF. Such an e-e correlation induced period increase may also be deduced by a thorough look at $\langle \hat{n}\rangle(x,t)$ during uv excitation of He in Fig.~\ref{fig.1dhedensity}~(b) and \ref{fig.1dhedensity}~(c). Moreover, Fig.~\ref{fig.dt}~(b) indicates how the evolution of the DM, including e-e correlations, depends on the initial state, cf.~the dashed
and dashed-dotted lines.
As expected, starting from the self-consistent initial state (TD2ndB$^1$) produces a DM which best matches the exact result. In contrast, the perturbation of the Hartree-Fock ground state (TD2ndB$^2$, equivalent to TD2ndB in Fig.~\ref{fig.dt}~(a)) produces a different time response in the DM amplitude. This, generally, lead [similar to the situation in Fig.~\ref{fig.1dhedensity}~(b)] to additional excitations which are not of dipole character.

\begin{figure}[t]
\includegraphics[width=0.48\textwidth]{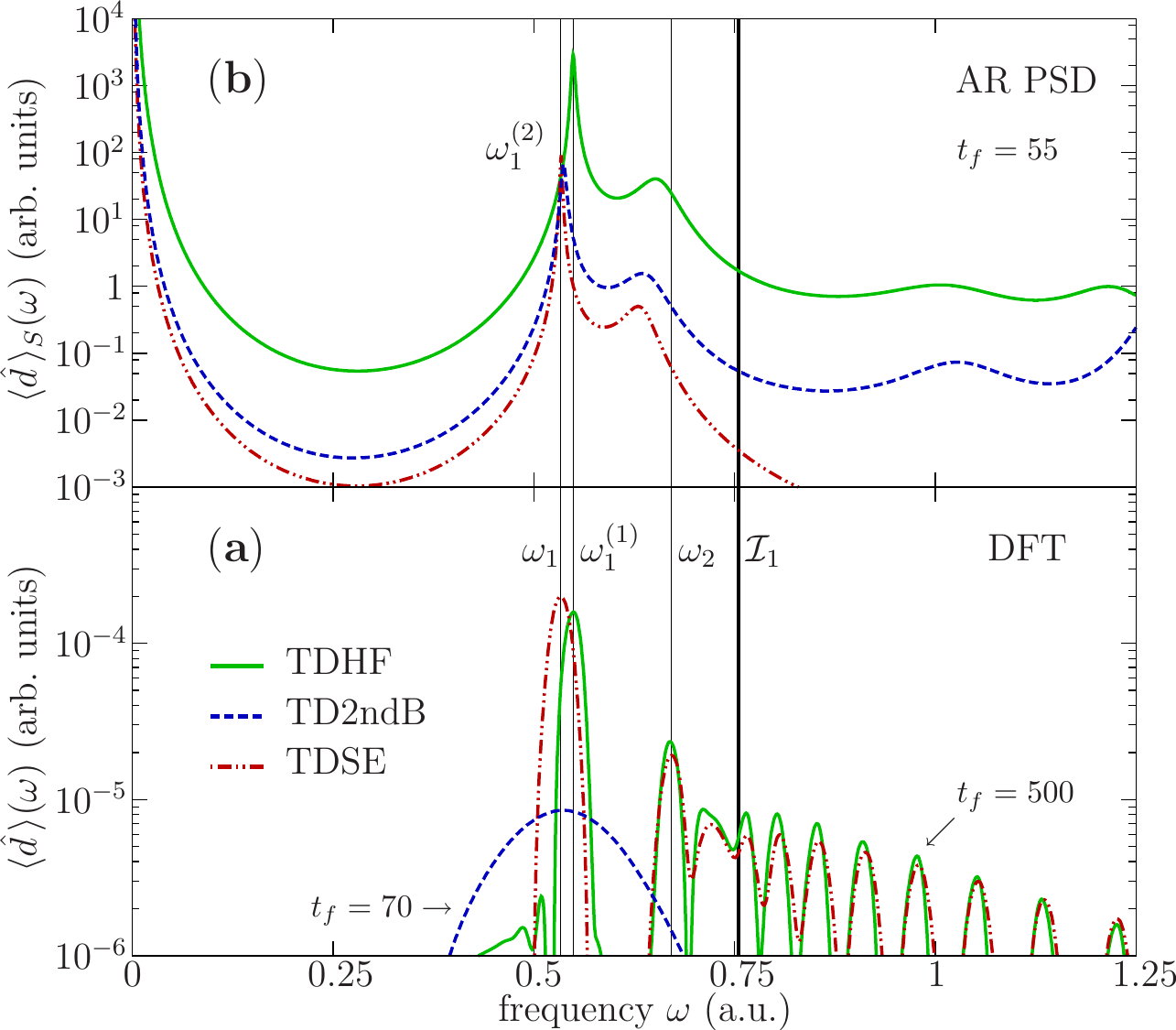}
 \caption{(Color online) Dipole strength $\langle\hat{d}\,\rangle(\omega)$ for 1D He. (a) Discrete Fourier transform with Hamming window from TDHF (solid) and TDSE (dash-dotted) time-series $\langle\hat{d}\,\rangle(t)$ of length $t_f=500$. The first excited state appears at $\omega_1=0.533$ (TDSE) and $\omega_1^{(1)}=0.549$ (TDHF), the second excited singlet state has frequency $\omega_2=0.671$. ${\cal I}_1=0.755$ denotes the first ionization threshold. Note, that peaks beyond ${\cal I}_1$ are artificial box states. (b) Auto-regression power spectral density $\langle\hat{d}\,\rangle_S(t)$ of short TDHF, TD2ndB, and TDSE time-series $\langle\hat{d}\,\rangle(t)$ of length $t_f=55$ computed from Eq.~(\ref{arpsd}) with model order $S=1000$. In TD2ndB approximation, the first excited state appears at frequency $\omega_1^{(2)}=0.537$.
}\label{fig.domega}
\end{figure}

A more detailed analysis of the DM is possible in terms of the dipole spectrum $\langle\hat{d}\,\rangle(\omega)\propto\int_0^{t_f}\!\dint{}{t}\,\mathrm{e}^{-\ii \omega t}\langle\hat{d}\,\rangle(t)$ with $t_f$ being the final propagation time. However, due to the two-time structure of the SKKBE~(\ref{skkbe}), typical TD2ndB calculations are limited to relatively short times of approximately $t_f\lesssim100$~a.u.~even if the parallel algorithm of Sec.~\ref{subsec:mpi} is being applied. Consequently, a discrete Fourier transform (DFT) reveals poor resolution, cf.~the dashed line in Fig.~\ref{fig.domega}~(a). In the case of TDHF (solid line) and TDSE (dashed-dotted line) much better resolution can be achieved for $\langle\hat{d}\,\rangle(\omega)$ such that, in Fig.~\ref{fig.domega}~(a), one recovers the series of one-electron excitations $\omega_i$ (excited singlet states of 1D He) which converges towards the first ionization threshold ${\cal I}_1=0.755$. The peaks beyond ${\cal I}_1$ indicate transitions to eigenstates 
of the finite simulation box. Overall, there is a shift concerning the first excited singlet state while energetically higher excitations seem to be less affected by e-e correlations. Concentrating on the first excited state of frequency $\omega_1$, we observe that it is overestimated by TDHF ($\omega_1^{(1)}=0.549$)---the exact value is $0.533$. Thus, it is interesting to analyze the behavior of TD2ndB ($\omega_1^{(2)}$). However, as usual Fourier transform of the DM time series is not well applicable, we have to rely on other methods which are not limited by the total length of the signal, e.g.~harmonic inversion (HI)~\cite{mandelshtam97}. Nevertheless, the best results have not been obtained by HI techniques, but by using an auto-regression (AR) spectral analysis~\cite{marple87} where an auto-correlation model of order $S$ is applied to the time-dependent dipole moment:
\begin{align}
\langle \hat{d}\,\rangle(t)&=\langle \hat{d}_0\rangle+\sum_{s=1}^{S}c_s \langle \hat{d}\rangle(t-s\Delta t)\,,&
\Delta t&=\frac{t_f}{S}\;,
\end{align}
with time average $\langle \hat{d}_0\rangle=\int_0^{t_f}\dint{}{\bar{t}}\,\langle \hat{d}\,\rangle(\bar{t})$ and AR coefficients $c_s$ which, e.g., follow from the Yule-Walker equations~\cite{marple87}. The AR coefficients then give direct access to the power spectral density (PSD) via
\begin{align}
\label{arpsd}
 \langle\hat{d}\,\rangle_S(\omega)\propto\left|1-\sum_{s=1}^{S}c_s \exp(-\ii s \omega\delta t/2)\right|^{-2}\;,
\end{align}
which is periodic in $\omega_{S}=4\pi \Delta t^{-1}=4\pi S/{t_f}$. Fig.~\ref{fig.domega}~(b) summarizes the AR PSD result for DM time-series of fixed length $t_f=55$~\cite{expl_arpsd}. Concerning the first excited state, TDHF (solid line) and TDSE (dashed-dot-dotted line) lead to well pronounced peaks which exactly equal the DFT result of Fig.~\ref{fig.domega}~(a). In the same manner, we are now able to explore the TD2ndB data and find $\omega_1^{(2)}=0.537$, cf.~the dashed line in Fig.~\ref{fig.domega}~(b). This frequency corresponds to a remarkable improvement of $75$\% of the deviation Hartree-Fock to exact. In addition, in $\langle\hat{d}\,\rangle_S(\omega)$, also the second excited singlet state is resolved. The general shift towards smaller frequencies $\omega_2$ is, thereby, caused by the shortness of the time series. In summary, the AR spectral analysis, being superior to DFT, allows for an appropriate characterization of short $\langle\hat{d}\,\rangle(t)$ time series. In particular, we conclude that, concerning dynamic properties of the 1D He model, TD2ndB shows correlation induced features to a substantial level, though, in the present analysis, it was not possible to show how well two-electron excitations~\cite{hochstuhl10_jpcs}---emerging between ${\cal I}_1$ and ${\cal I}_2=2.2383$ and being absent in TDHF---are resolved in TD2ndB approximation. This lack is only due to the limited propagation time, and, indeed, we expect these features of e-e correlations to be included on a similar level.

\begin{figure}[t]
\includegraphics[width=0.48\textwidth]{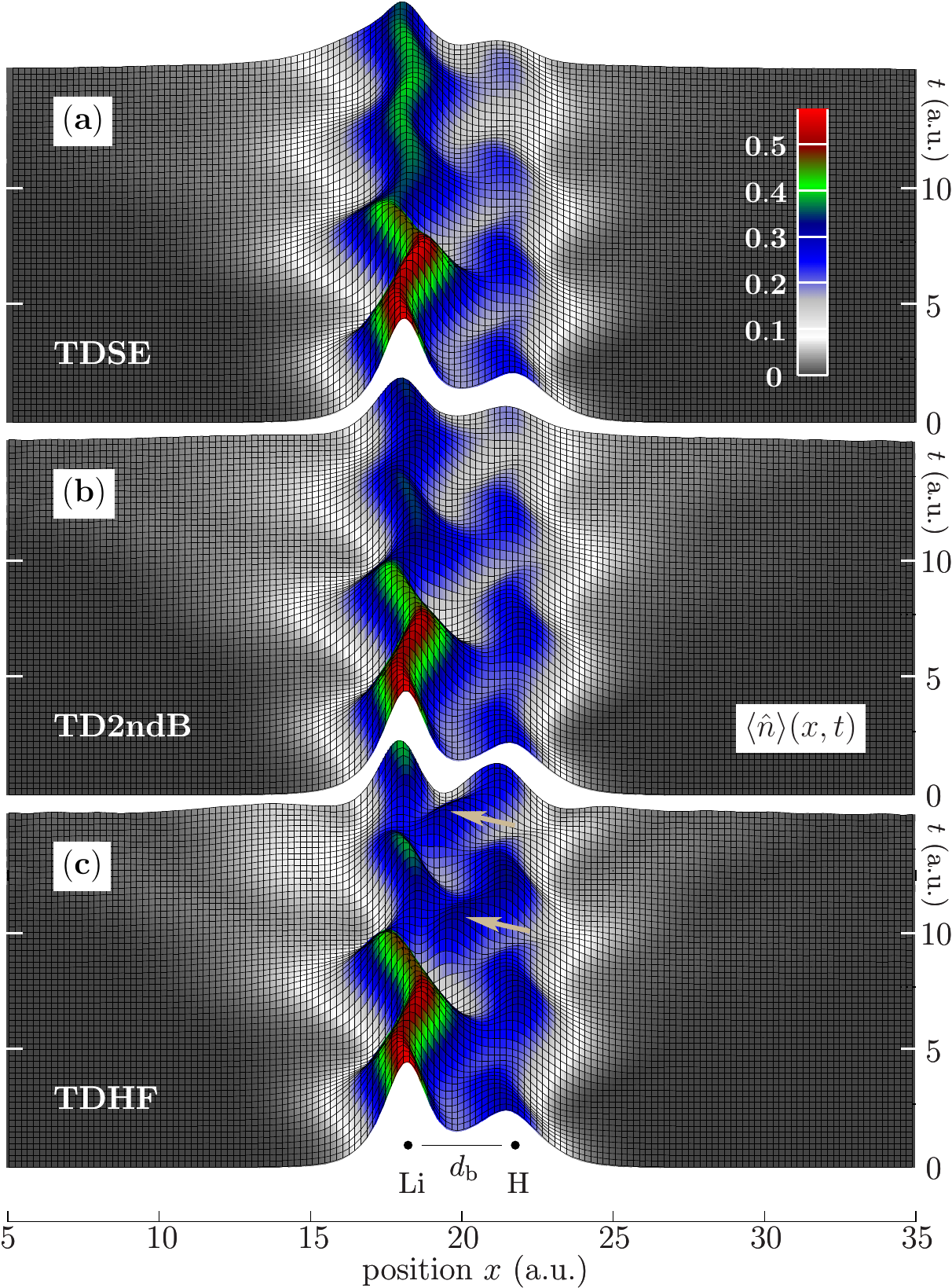}
 \caption{(Color online) Time-dependent one-electron density $\langle\hat{n}\rangle(x,t)$ (plotted nonlogarithmically) for the response of LiH to an xuv field with intensity $E_0=0.75$ and frequency $\omega_0=1.3$. While panel (a) indicates the [exact] TDSE result, panel (b) and (c), respectively, refer to self-consistent TD2ndB and TDHF calculations. Arrows in (c) indicate particularly strong deviations of TDHF compared to TDSE.
}\label{fig.1dlihdensity}
\end{figure}

\begin{figure}[t]
\includegraphics[width=0.48\textwidth]{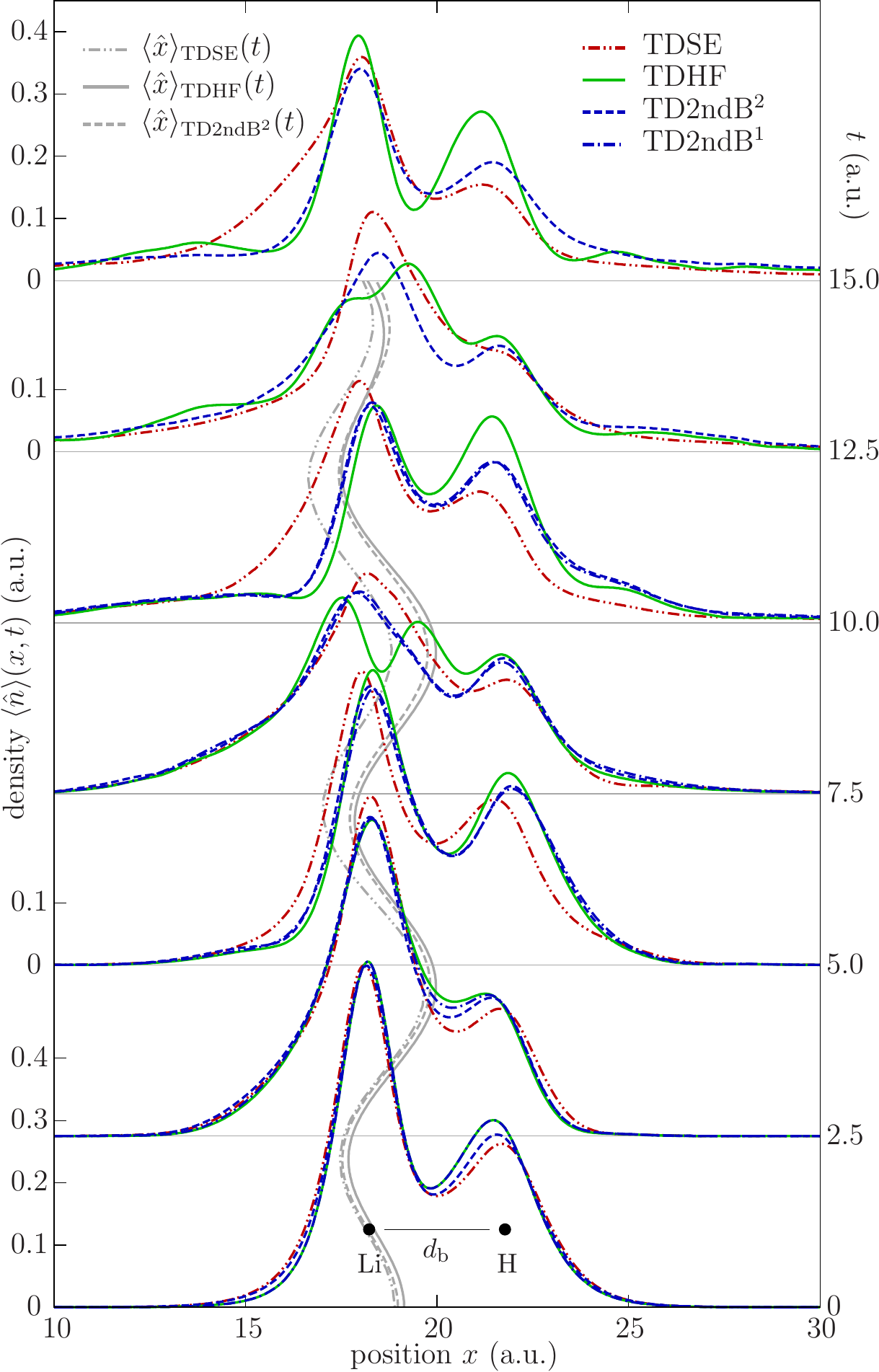}
 \caption{(Color online) One-electron density $\langle\hat{n}\rangle(x,t)$ at different times $t$ for the response of LiH to an xuv field (with parameters as in Fig.~\ref{fig.1dlihdensity}). The TDHF (solid) and TD2ndB$^{1}$ result (dashed) include self-consistent initial states (with different but fixed $d_\mathrm{b}$) while TD2ndB$^2$ (dotted) starts from the Hartree-Fock ground state [only shown for $t\leq10$)]. Note, that the self-consistent solution of the TDSE (dashed-dot-dotted) is not fully accurate due to lack of spatial resolution---compare with the exact ground state (dashed-dotted line) for $t=0$. Moreover, the gray curves show the time-dependent expectation value $\langle\hat{x}\rangle(t)$ of the electron position.
}\label{fig.1dlihdensity2}
\end{figure}

\subsection{\label{subsec:lih}Lithium hydride}
Using two-time NEGFs, it is possible to describe even more complex systems than He in nonequilibrium. As a proof of principle, we consider the response of the neutral molecule lithium hydride (LiH) to an xuv field of intensity $E_0=0.75$ ($2.0\cdot10^{16}$ W/cm$^2$) and frequency $\omega=1.3$ (photon energy $35.5$ eV). LiH is a heteronuclear molecule ($M=2$, $Z_1=3$, and $Z_2=1$ in Eq.~(\ref{tdp})) including four electrons and complies with the restricted NEGF ansatz~(\ref{gdef}) when prepared as singlet state, i.e.~with closed-shell spin configuration. Prior to excitation, we have fixed the Li-H bond lengths $d_\mathrm{b}=x_{0,2}-x_{0,1}$ [in terms of the Born-Oppenheimer separation of electronic and nuclear motion] to the self-consistent values as obtained by scanning the potential energy surface~\cite{balzer10_pra}: $d_\mathrm{b}=3.386$ (Hartree-Fock), $d_\mathrm{b}=3.505$ (second Born), $d_\mathrm{b}\approx3.6$ (exact). The resulting ground state densities $\langle\hat{n}\rangle(x,0)$ are indicated in Fig.~\ref{fig.1dlihdensity2}, cf.~the lines at the bottom of the plot ($t=0$). Computing the LiH time evolution from the TDSE in four spatial coordinates is, in contrast to helium, already ambitious but nevertheless feasible. In comparison, going from 1D He to the LiH model solving the SKKBE for the one-particle NEGF is conceptually and computationally very simple as, with the increase of particle number, only the normalization of the Green's function is concerned. Fig.~\ref{fig.1dlihdensity} reveals the picture of the xuv induced electron dynamics over more than three laser cycles as obtained from TDSE (top), time-dependent [self-consistent] second Born approximation (center) and time-dependent Hartree-Fock approximation (bottom). Thereby, we have used an FE-DVR basis of size $n_b=71$ and a simulation box of $x_0=40$. As general trend, we observe, in $\langle\hat{n}\rangle(x,t)$, that TD2ndB is considerably superior to TDHF. In particular, the time-dependent density between the Li and the H atom behaves completely different than TDSE in the case of TDHF, see e.g.~the density around $x=20$ for $t=9$ and $t=13$ indicated by arrows. This is essentially corrected in TD2ndB approximation.

For specific times $t$, Fig.~\ref{fig.1dlihdensity2} allows for a direct and more detailed comparison of the exact and approximate one-electron densities close to the atomic nuclei. As in Fig.~\ref{fig.dt}, TD2ndB$^1$ (dashed-dotted line) means that the initial state is uncorrelated [thus on the Hartree-Fock level], while TD2ndB$^2$ (dashed line) refers to self-consistent ground state preparation. Following the time-evolution, TD2ndB$^2$ leads to the correct changes towards the TDSE solution including strong depopulation of electrons around the hydrogen atom for $t>10$. Also, the TD2ndB$^2$ density is more smooth whereas TDHF shows additional maxima being absent in, both, TD2ndB approximation and the exact result (dashed-dot-dotted line), see e.g.~the density for $t=7.5$ at $x\approx20$. In addition, we observe that starting from the correlated (TD2ndB$^2$) or uncorrelated ground state (TD2ndB$^1$) does not make much density difference for later propagation times $t\gtrsim5$. Averaging over coordinate space allows us, moreover, to trace the time-dependent expectation value of the electron position $\langle\hat{x}\rangle(t)$  which, being initially close to the Li nucleus, starts to oscillate with the xuv field, cf.~the gray lines in Fig.~\ref{fig.1dlihdensity2}. Thereby, TD2ndB, again, gives consistent results, although, for later times, deviations to TDSE increase. To this end, we note that the solution of the full four-particle TDSE could have been performed only with restricted spatial resolution whereas the NEGF results are converged. Thus, comparisons should be drawn carefully. Nevertheless, for confirming the main trends [when e-e correlations are being included in the NEGF calculation], the TDSE result is fully adequate.

\section{\label{sec:conclusion}Conclusion}
The (x)uv laser-induced, short-time electron dynamics in the 1D helium and lithium hydride model have been investigated using a NEGF approach. To this end an efficient grid-based approach of Ref.~\cite{balzer10_pra} was extended to nonequilibrium situations. Starting from initial states with different degrees of e-e correlation and self-consistency, it has been shown that NEGFs are well applicable to resolve the time dependence and that dynamic correlation effects are approximated to a substantial and valuable level already in the case of TD2ndB.
This is obtained from (i) analyzing the few-cycle (one-electron) density response of He and LiH to strong laser fields and (ii) from exploring spectral properties such as the singly-excited spectrum of 1D He, cf.~Sec.~\ref{sec:results}. Overall, results at TD2ndB level are
found to be closer to exact solutions of the TDSE than to TDHF. Thereby, the present restriction of NEGF calculations to examples of two- and four-electron systems stems from the availability of exact reference data which allows for a proof of principle. Exact solutions of the TDSE for $N\gtrsim4$ are essentially at the limit concerning spatial resolution whereas a solution of the SKKBEs is not explicitly sensitive to the particle number. This potential of direct extension to larger model atoms or molecules [where exact calculations are impossible] affirms the power of the present NEGF approach.

As the TD2ndB results show important time-dependent features, we believe, that the present analysis will stimulate and support further investigations of strong-field laser-atom interactions based on the two-time NEGF approach. Moreover, due to the distributed memory concept (Sec.~\ref{subsec:mpi}) which is capable to organize the huge demand of RAM involved in the memory kernel, we are now able to perform the two-time propagation of the SKKBEs in an optimized and efficient way. This issue, in future work, will enable more extended calculations with increased FE-DVR basis size ($n_b$) and longer propagation times ($t_f$) and, hence, fortifies a quantum kinetic approach to atoms and molecules and their interaction with (strong) laser fields.

\begin{acknowledgments}
This work was supported by computing time at the North German Super Computer Alliance (HLRN) via grant shp0006, the Leibniz Rechenzentrum der Bayerischen Akademie der Wissenschaften (HLRB) and by the Deutsche Forschungsgemeinschaft via grant BO1366/9-1.
\end{acknowledgments}

\appendix*
\section{Self-energies}
For a complete discussion
of the matrix equation (\ref{skkbe}) in Sec.~\ref{subsec:eom}, we, here, give the definitions of the self-energies: (i) In time-dependent Hartree-Fock approximation (TDHF), the self-energy is given by $\Sigma_\xi[g,u]=\delta_{\cal C}(t-t')\Sigma^{\mathrm{HF}}_\xi[g,u]$ with
\begin{align}
&\Sigma^{\mathrm{HF},i_1i_2}_{\xi,m_1m_2}[g,u](t,t')&\nonumber\\
=&-\mathrm{i}\,\xi\,\delta^{i_1i_2}_{m_1m_2}\sum_{i_3m_3}\tilde{u}_{m_1m_3}^{i_1i_3}\,g_{m_3m_3}^{i_3i_3}(t,t^+)&\nonumber\\
&+\mathrm{i}\,\tilde{u}_{m_2m_1}^{i_2i_1}\,g_{m_2m_1}^{i_2i_1}(t,t^+)\;,\nonumber
\end{align}
(ii) in time-dependent second Born approximation (TD2ndB), it is $\Sigma_\xi[g,u]=\delta_{\cal C}(t-t')\Sigma^{\mathrm{HF}}_\xi[g,u]+\Sigma^{\mathrm{2ndB}}_\xi[g,u]$, where
\begin{align}
&\Sigma^{\mathrm{2ndB},i_1i_2}_{m_1m_2}[g,u](t,t')&\nonumber\\
=&\sum_{i_3m_3}\sum_{i_4m_4}\tilde{u}_{m_1m_4}^{i_1i_4}\,\tilde{u}_{m_2m_3}^{i_2i_3}\left\{\xi\,g^{i_1i_2}_{m_1m_2}(t,t')\,g^{i_4i_3}_{m_4m_3}(t,t')\right.&\nonumber\\
&\left.-g^{i_1i_3}_{m_1m_3}(t,t')\,g^{i_4i_2}_{m_4m_2}(t,t')\right\}g^{i_3i_4}_{m_3m_4}(t',t)\;.\nonumber
\end{align}



\begin{thebibliography}{100}
\bibitem{expl_arpsd} We note, that the initial state was taken on the Hartree-Fock level, cf.~Fig.~\ref{fig.dt}, thus, it is not self-consistent. However, we expect this to have only minor effects on the considered dipole transitions.
\end{thebibliography}
\end{document}